\documentclass[useAMS]{mn2e}
\usepackage{supertabular}
\usepackage{longtable}
\usepackage{graphicx}

\title [C-Miras, Velocities and Distances] {Carbon-Rich Mira Variables: 
Radial Velocities and Distances}
\author [Menzies, Feast \& Whitelock] {John W. Menzies$^{1}$,
 Michael W. Feast$^{3}$, Patricia A. Whitelock$^{1,2,3}$\\
$^{1}$ South  African Astronomical Observatory, PO Box 9, 7935 Observatory,
South Africa, jwm@saao.ac.za\\
$^{2}$ National Astrophysics and Space Science Programme, University of
Cape Town, 7701, Rondebosch, South Africa\\
$^{3}$ Astronomy Dept., University of Cape Town, Rondebosch, 7701, 
South Africa\\
}
\begin{document}
\maketitle
\begin{abstract}
Optical radial velocities have been measured for 38 C-type Mira variables.
These data together with others in the literature are used to study the
differences between optical and CO mm observations for C-Miras and the necessary
corrections to the optical velocities are derived in order to obtain
the true radial velocities of the variables. The difference between
absorption and emission line velocities is also examined. 
A particularly large difference ($+30\, \rm km\,s^{-1}$) is found in the case of
the $H\alpha$ emission line.
A catalogue is given
of 177 C-Miras with estimated distances and radial velocities. The distances
are based on bolometric magnitudes derived in Paper I using SAAO
observations or (for 60 of the stars) using non-SAAO photometry. In
the latter case the necessary transformations to the SAAO system are derived.
These data will be used in paper III to study the kinematics of the C-Miras.
\end{abstract}

\section {Introduction}
This is the second paper of three reporting a study of the nature of
Galactic carbon-rich Mira variables. An important part of this study will
be the analysis of the galactic kinematics of C-Miras. Although 
there have been extensive radial
velocity studies of oxygen-rich Miras both at optical and radio
frequencies (see the tabulation  in Feast \& Whitelock 2000), the 
radial velocity work
on C-Miras has been much less extensive. 
For instance, the velocity dispersion of these objects
derived by Feast (1989) depended on radial velocities of only 36 stars.
The early work is summarized in the catalogue of Sanford (1944). This,
like several other radial velocity studies deals with both variable
and non-variable carbon stars. Amongst the papers containing optical
radial velocities of C-Miras are those of Smak \& Preston (1965),
Yamashita (1974), Dean (1976), Walker (1979), Aaronson et al. (1989, 1990),
and those of Sanford (e.g. 1950) and Barnbaum (1992ab, 1994) which 
are higher resolution studies. 

In the present paper we report the optical radial velocities of 38 C-Miras
including 18 with no previous published velocity. The stars were selected
from previously known C-Miras without known radial velocities and from stars
in the $JHKL$ programme discussed in Whitelock et al (2005) (paper I). For
comparison purposes there was also some overlap with C-Miras having
previously measured radial velocity. This overlap includes some stars with
radial velocity measures in Aaronson et al. (1989) which have been shown in
paper I to be Mira variables. There are also now extensive lists of radial
velocities of stars from CO mm observations, most of which are tabulated in
Olofsson et al. (1993), Loup et al. (1993), Groenewegen et al. (1999) or
Groenewegen et al. (2002). These lists contain velocities for some known
C-Miras as well as a considerable number that were shown to be Miras in
paper I.  The considerable overlap between optical and CO velocities now
enables us to carry out an improved discussion of the correction necessary
to put the optical velocities on the CO system which is believed to
represent the true radial velocity of the star. Not only is this correction
needed in the present work but it should be useful in the future since it is
now possible to derive optical velocities for C-Miras that are too faint for
CO velocity measurements (e.g. those in the Magellanic Clouds and some more
distant systems). The optical/CO comparison also yields an estimate of the
uncertainty of a typical optical velocity due both to stellar pulsation and
observational error. This is important when estimates are being made of the
velocity dispersion in a group of C-Miras. Note that a sub-sample of the
stars chosen for study as possible Miras in paper I were selected by similar
criteria, based on IRAS colours, as were the stars measured for CO velocity
in Groenewegen et al. (2002). {\bf We also discuss the corrections necessary
to place $JHKL$ observations of C-Miras by others on the SAAO system and use
the corrected data to derive bolometric magnitudes of these stars.

\section{Observations} 

All SAAO spectra were obtained with the Unit Spectrograph attached to the
1.9-m telescope at Sutherland. A SITe 1798x266 pixel CCD with 13 $\mu$m
pixels was used in the spectrograph camera. An 830 line/mm grating was used
in first order with a 300~$\mu$m slit to give spectra covering the
range, 5600 to 6860 \AA\, at a resolution of about 2 \AA. Comparison Cu/Ar
arc spectra were obtained before and after each object spectrum. Several
different Carbon stars were observed as standard radial velocity stars each
night. These standards were chosen from a list of stars used by 
Walker (1979) in
an earlier extensive radial velocity programme.

A mean template was constructed from the spectra of all the standards
shifted to zero velocity, and this was cross-correlated with the
programme-star spectra as well as with the individual spectra. The resulting
mean velocities for the standards are given in Table 1. 
This table gives, the name of the object, the velocity from
the catalogue of Wilson (1953), the velocity from Walker (1979),
our mean velocity, the dispersion amongst our measures, the
difference of our mean from that of Wilson, the number of our
spectra and the spectral type from the literature.
Comparison
with the catalogue velocities shows that they are well-reproduced by our
measurements.

A number of the programme stars and two of the standards showed H$\alpha$
emission lines. Gaussian profiles were fitted to these and the central
wavelengths compared to the assumed rest wavelength, 6562.82 \AA\, of
H$\alpha$. It should be noted that these velocities depend on how well the
positions of the comparison arc lines, which are unfortunately sparsely
distributed in the region around H$\alpha$, were fitted in each individual
case. Only the standard, V Oph, was observed on more than one night, but the
two velocities obtained agree to 2 $\rm km\,s^{-1}$. The cross-correlation
velocities only depend on how well in the mean all arc line positions in the
wavelength range used were fitted.

The individual results for all programme objects are given in
Table 2.

\begin{table}
\caption{Standard Stars: Heliocentric Radial Velocities.}\label{std}
\begin{tabular}{rrrrrrcl} 
\hline
Object  & \multicolumn{2}{c}{V($\rm km\,s^{-1}$)} & Obs  & \multicolumn{1}{c}{$\sigma$}& C-O& N& Sp-Type\\
        & \multicolumn{1}{c}{Cat}&\multicolumn{1}{c}{W}&&&&&\\
\hline
HD75021   &   11.0& 14.0  &  12.1 &  7.2  &  --1.1  &    6  &  C (R8) \\
BN Mon    &   27.7& 25.9  &  24.4 &  6.8  &   3.3  &    7  &  N0-2 \\
R Lep     &   32.4& 32.1  &  33.1 &   -   &  --0.7  &    1  &  C7,6E \\
U Hya     &  --25.0&--24.3  & --16.9 &  3.0  &  --8.1  &    7  &  C \\
V Oph     &  --37.4&--39.7  & --45.2 &  3.6  &   7.8  &    4  &  C \\
X Tra     &   --3.7& --1.8  &   4.1 &  3.7  &  --7.8  &    8  &  C+ \\
Y Hya     &    3.0&  5.7  &   1.4 &  5.0  &   1.6  &    6  &  C \\
\hline
\end{tabular}
\end{table}

\begin{table}
\caption{New Heliocentric Radial Velocities.}\label{results}

\begin{tabular}{lrrc}
\hline
Object      &      V(abs)  &     V(em)  &  JD Obs.  \\
            &      \multicolumn{2}{c}{($\rm km\,s^{-1}$)}    & 2450000+ \\
\hline


V718 Tau  &  4.2  &  --29.2  &  514.3 \\
V718 Tau  &  --5.9   &  --36.2  &  516.3 \\
QS Ori  &  44.5  &  8.8  &  514.3 \\
QS Ori  &  42.5   &  9.8  &  516.3 \\
V617 Mon  &  22.2  &  --15.80  &  514.3 \\
V617 Mon  &  28.9  &  --10.1  &  516.3 \\
ZZ Gem  &  76.9  &  35.8  &  516.3 \\
V503 Mon  &  --25.9  &  --47.3  &  514.3 \\
V503 Mon  &  --0.3  &  --24.9  &  516.3 \\
RT Gem  &  98.3   &  77.6  &  514.3 \\
RT Gem  &  133.5   &  107.8   &  516.3 \\
CG Mon  &  21.2   &    &  514.3 \\
CG Mon  &  31.2  &    &  516.3 \\
CL Mon  &  30.9  &  4.6  &  514.4 \\
CL Mon  &  32.6  &  9.6   &  514.4 \\
CL Mon  &  37.8  &  13.0  &  516.3 \\
R CMi  &  25.2   &    &   513.3 \\
R CMi  &  35.0   &    &  514.4 \\
R CMi  &  37.0   &    &  516.4 \\
07097--1011  &  91.3   &  47.0  &  515.3 \\
VX Gem  &  43.6  &    &  513.3 \\
07136--1512  &  81.6   &    &   515.3\\
07136--1512  &  88.7   &    &  517.3 \\
07217--1246  &  39.9  &  --2.3  &  515.3 \\
07223--1553  &  76.8  &    &  515.3 \\
07223--1553  &  79.3   &    &  517.3 \\
07293--1832  &  109.7   &    &   515.3\\
$[$W71b$]$007--02  &  74.5  &  46.1  &  515.4 \\
$[$W71b$]$007--02  &  76.4   &  46.5   &  517.3 \\
$[$W71b$]$008--04  &  102.0   &  81.6 &  515.4 \\
V831 Mon  &  30.5   &  --16.5  &  517.3 \\
FF Pup  &  65.6   &  38.1  &  513.4 \\
$[$W71b$]$021--06  &  12.8   &  --19.7  &  515.4 \\
$[$W71b$]$021--06  &  26.8  &  --16.7   &  517.4 \\
$[$W71b$]$029--02  &  80.7  &  45.2   &  517.4 \\
IQ Hya  &  15.0  &  --17.6  &  517.4 \\
$[$W71b$]$046--02  &  42.7   &  14.0   &  515.4 \\
$[$ABC89$]$Vel 44  &  62.2  &  28.6  &  517.4 \\
$[$ABC89$]$Pup 42  &  125.1  &    &  517.3 \\
$[$ABC89$]$Ppx 19  &  57.2  &  24.7  &  517.4 \\
$[$ABC89$]$Ppx 40  &  72.3  &  41.47  &  515.4 \\
$[$ABC89$]$Ppx 40  &  92.0   &  63.7  &  517.4 \\
R Pyx  &  47.1   &  12.1   &  513.4 \\
CZ Hya  &  16.7   &  0.3   &  513.4 \\
CZ Hya  &  22.0  &  --8.6   &  514.5 \\
CZ Hya  &  22.2   &  0.1  &  514.5 \\
CZ Hya  &  31.6   &  9.9   &  516.4 \\
TV Vel  &  9.6   &  --12.4   &  512.4 \\
TV Vel  &  6.1   &  --20.4  & 514.5  \\
FU Car  &  19.1  &  --12.3  &  512.5 \\
$[$ABC89$]$Car 10  &    &  --14.6   &  515.5 \\
$[$ABC89$]$Car 10  &    &  8.5  &  517.5 \\
$[$ABC89$]$Car 28  &  78.6   &  32.3   &  517.5 \\
$[$ABC89$]$Car 32  &  11.2  &    &  517.5 \\
$[$ABC89$]$Car 73  &  15.5  &    &  517.5 \\
$[$ABC89$]$Car 87  &  21.7  &  --11.8   &  515.5 \\
$[$ABC89$]$Car 87  &  40.7   &  7.8   &  517.5 \\
$[$W65$]$c1  &  20.4  &    &  513.5 \\
$[$W65$]$c2  &  --1.1   &    &  513.5 \\
$[$ABC89$]$Cen 4  &  0.9   &    &  513.5 \\
$[$ABC89$]$Cen 78  &  27.9  &    &  515.5 \\
\multicolumn{4}{l}{\it continued in the next column...}\\
\hline
\end{tabular}
\end{table}
\begin{table}
\setcounter{table}{1}
\caption{continued...}
\begin{tabular}{lrrc}
\hline
Object      &      V(abs)  &     V(em)  &  JD Obs.  \\
            &      \multicolumn{2}{c}{($\rm km\,s^{-1}$)}    & 2450000+ \\
\hline
V Cru  &  --31.2  &  --50.2   &  512.6 \\
$[$ABC89$]$Cir 1  &  --42.6   &    &  515.5 \\
$[$ABC89$]$Cir 27  &  --30.7  &  --57.6   &  515.6 \\
14395--5656  &  --70.5   &  --99.8  &  515.6 \\
NP Her  &  1.7  &  --26.9   &  512.6 \\
SZ Ara  &  --7.7   &  --30.8  &  513.6 \\
V617 Sco  &  --8.6   &  --26.6  &  513.6 \\
\hline
\end{tabular}

\end{table}

\section{Velocity Comparisons}
  Optical radial velocities are generally published in a heliocentric system,
as are the new velocities given in this paper (Table 2). The mm
CO velocities are published in a system referred to as local standard of
rest (lsr). Despite the fact that there is still not general agreement on the
motion of the Sun with respect to the circular velocity of galactic rotation
at the solar position (see, e.g. Feast 2000), the value adopted by the CO
observers for this quantity is rarely stated. However it appears that
they generally adopt a solar motion of 20 $\rm{km\,s^{-1}}$ towards $\alpha
= 18^{h}$ and
$\delta = +30^{\circ}$ (1900). We have therefore converted  CO velocities
to heliocentric using these values.
 
  It is generally believed that the CO velocities, being derived from a region
outside the main pulsating stellar atmosphere, are a good measure
of the systemic velocity of the C-Mira. Barnbaum (1992) who carried out
high resolution optical work 
on carbon stars found an offset between her absorption-line
velocities and the CO velocities. Her data for C-Miras gives:\\

\begin{equation}
\rm{CO} - \rm{Optical} = -5.8 \pm 1.5 \ \  \rm{km\,s^{-1}} 
\end{equation}
for 10 stars with a mean period of 443 days and with a standard deviation of
$4.7\, \rm{km\,s^{-1}}$. Barnbaum (1992) notes that the CO measures are
generally consistent to about $1\, \rm{km\,s^{-1}}$. Thus her results indicate
that her mean velocities have uncertainties of about $5\, \rm{km\,s^{-1}}$. In
the currently available sample of C-Miras and using straight means of all
the optical measures for each star one finds:\\
\begin{equation}
\rm{CO} - \rm{Optical} = -3.8 \pm 1.1 \ \ \rm{km\,s^{-1}}
\end{equation}
for 23 stars with a mean period of 444 days and a standard deviation
of $5.1\, \rm km\,s^{-1}$. In Feast \& Whitelock (2000) it was found that
the offset of the maser OH velocity from the optical velocity  for
O-Miras was given by:\\
\begin{equation}
\rm{OH} - \rm{Optical} = -0.015P + 1.31 \ \ \rm {km\,s^{-1}}.
\end{equation}
This gives $-5.3\, \rm{km\,s^{-1}}$ at a period of 443 days, an offset similar
to the above values for the CO/Optical difference of C-Miras. 
In the above and following calculations the optical radial velocity of QS Ori
published by Dean (1976) ($+9\,\rm{km\,s^{-1}}$) has been omitted since in
the present paper we find (Table 2) $+43\,\rm {km\,s^{-1}}$ and the 
heliocentric CO
velocity is $+38\, \rm{km\,s^{-1}}$.
If the sample of 23 C-Miras just discussed} is divided by period one finds:\\
\begin{equation}
\rm{CO} - \rm{Optical} = -4.6 \pm 1.4 \ \ \rm{km\,s^{-1}}
\end{equation}
for 11 stars with a mean period of 385 days and a standard deviation of
$\rm 4.4\, km\,s^{-1}$, and\\
\begin{equation}
\rm{CO} - \rm{Optical} = -2.2 \pm 1.9 \ \ \rm{km\,s^{-1}}
\end{equation}
for 12 stars with a mean period of 498 days and a standard deviation of
$\rm {6.8 \,km\,s^{-1}}$. At these two periods, equation 3 predicts
$-4.5$ and $-6.2\, \rm{km\,s^{-1}}$. Although the results for the two period
groups of C-Miras do not differ significantly they suggest that
the CO -- Optical difference  is
not the same for C-Miras and O-Miras at the longer periods.
In view of the above we have applied a correction of
$-4\, \rm{km\,s^{-1}}$ to all the optical velocities we use in our kinematic
analysis.

Barnbaum \& Hinkle (1995) suggest that the radial velocities of
C-Miras derived in the near infrared, particularly from lines of the
IR CN bands, are close to the CO velocities. Leaving out from their
sample UV Aur (Symbiotic), V Hya (a peculiar binary) and SS Vir (not
a normal AGB star according to Barnbaum \& Hinkle and classed as 
a semiregular variable, not a Mira (Whitelock et al. paper I)), we find:\\
weighted per star
\begin{equation}
\rm{CO} - \rm{CN(IR)} = -1.9 \pm 2.4 \ \  \rm{km\,s^{-1}}
\end{equation}
for 9 stars, standard deviation $7.2\, \rm {km\,s^{-1}}$.
Or, weighting per spectrum (the method adopted by Barnbaum \& Hinkle)
\begin{equation}
\rm{CO} - \rm{CN(IR)} = -0.9 \pm 1.5 \ \ \rm{km\,s^{-1}}
\end{equation}
for 15 spectra with a standard deviation of $5.7\, \rm{km\,s^{-1}}$. 
These results are not as clear cut as one might like, but we have chosen
to adopt the IR velocities without correction where they are used in
the kinematic analysis.  

\section{Non-SAAO Photometry}
In order to derive bolometric magnitudes we need near infrared
photometry as well as IRAS/MSX data in the mid-IR. 
Where possible we have used the SAAO data tabulated and discussed in paper I.
For other C-Miras with known periods, radial velocities 
and mid-infrared data we have taken near infrared observations from
the literature.
Because C-Miras
have different energy distributions in the near-IR to photometric
standard stars, it is necessary to consider carefully the transformations
between different photometric systems. The following sets of data
are useful for the present purpose and satisfactory transformations
to the SAAO system can be derived for them:
Jones et al. (1990), Noguchi et al. (1981),
Taranova \& Shenavrin (2004)(TS), ESO (Epchtein et al. 1990, Le Berte 1992,
Fouqu\'{e} et al. 1992, Guglielmo et al. 1993) 
and 2MASS (Cutri et al. 2003). Note that only
in the case of Le Bertre (1992) (ESO), 
TS and, to a lesser extent Jones et al. (1990)
were extensive observations of any one star available. Much of
the data are single epoch observations only. We do not need to use the
(mostly single epoch) observations of Aaronson et al. (1989) since
we have SAAO observations of all the stars of interest in their sample.

All the differences given in this section are in the sense SAAO minus Other.

\begin{table*}
\begin{center}
\caption[Non-SAAO photometry converted to the SAAO System.]
{Non-SAAO photometry converted to SAAO System.}\label{phot}
\begin{tabular}{llrrrrccrcclc}  
\hline
 name & IRAS & \multicolumn{1}{c}{$J$}& \multicolumn{1}{c}{$H$}& 
\multicolumn{1}{c}{$K$}& \multicolumn{1}{c}{$L$}& \multicolumn{1}{c}{A$_V$}
& \multicolumn{1}{c}{$m_{bol}$}& No. & P & source$^\dagger$\ &
source$^\ddagger$ & MSX \\ 
& & \multicolumn{6}{c}{(mag)} & & (day) & P & near-IR\\
\hline
V668 Cas  & 00247+6922 &      &      &  4.52&  1.93& 0.63 &  5.70 & 10& 650& J & J  &   \\
W Cas     & 00519+5817 &  4.64&  3.45&  2.83&  2.59& 0.67 &  6.19 & 14& 399& H & T  & X \\ 
HV Cas    & 01080+5327 &  5.32&  3.53&  2.23&  1.54& 0.55 &  5.83 &  2& 527&   & NM &   \\
PT Cas    & 01348+5543 &  7.07&  5.69&  5.12&      & 0.75 &  8.58 &  1& 300&   & M  &   \\
X Cas     & 01531+5900 &  4.72&  3.27&  2.42&  1.71& 0.64 &  5.90 &  4& 420& H & TNM& X \\
V596 Per  & 02293+5748 &      & 15.81& 11.09&      & 1.06 &  6.47 &  1& 815& G & M  & X \\
V409 Per  & 03063+4546 &  5.27&  3.89&  3.22&      & 0.50 &  6.75 &  1& 355&   & M  &   \\
V701 Cas  & 03186+7016 & 10.80&  7.36&  4.80&      & 0.52 &  6.05 &  1& 567& J & M  &   \\
KX Cam    & 03192+5642 &  8.64&  6.25&  4.68&      & 1.73 &  7.58 &  1& 430& N & M  & X \\
V384 Per  & 03229+4721 &      &      &  1.35&--0.37& 0.36 &  4.27 & 13& 537& O & J  &   \\
KY Cam    & 03238+6034 & 10.54&  7.36&  5.08&      & 1.10 &  6.88 &  1& 477& N & M  &   \\
Y Per     & 03242+4400 &  5.14&  4.21&  3.68&      & 0.49 &  6.82 &  1& 238& H & M  &   \\
V414 Per  & 03488+3943 &      &      &  2.13&  0.73& 0.46 &  5.36 & 11& 515& J & J  &   \\
V676 Per  & 03557+4404 & 13.22&  9.61&  6.96&      & 0.74 &  7.77 &  1& 482& N & M  &   \\
AU Aur    & 04504+4949 &  5.23&  3.85&  2.89&      & 0.67 &  6.35 &  1& 377& H & M  &   \\
DY Aur    & 05149+3511 &  6.53&  4.68&  3.23&      & 0.82 &  6.41 &  1& 474&   & M  & X \\
OP Aur    & 05247+3427 &  7.15&  5.59&  4.71&      & 1.42 &  8.09 &  1& 500&   & M  & X \\
V370 Aur  & 05405+3240 &      &      &  4.52&  2.03& 0.82 &  5.99 &  6& 732& C & J  & X \\
V393 Aur  & 05440+4311 &      &      &  3.37&  1.93& 0.57 &  6.50 &  6& 517& J & J  &   \\
AZ Aur    & 05576+3940 &  5.26&  3.88&  2.86&  2.07& 0.54 &  6.28 &  2& 416&   & NM &   \\
V1259 Ori & 06012+0726 &      & 10.28&  6.92&  2.43& 0.60 &  5.65 & 22& 696& B & E  &   \\
V Aur     & 06202+4743 &  5.37&  3.92&  3.02&      & 0.34 &  6.56 &  1& 349& H & M  &   \\
V713 Mon  & 06230--0930 & 8.92&  6.46&  4.60&  2.29& 0.65 &  6.75 & 26& 494& B & E  &   \\
V688 Mon  & 06342+0328 & 10.25&  6.97&  4.50&  1.55& 0.90 &  6.01 & 23& 653& B & E  & X \\
HX CMa    & 07098--2012 & 7.98&  5.52&  3.72&  1.44& 0.74 &  5.74 & 31& 725& B & E  &   \\
T Lyn     & 08195+3340 &  5.26&  3.92&  3.03&  2.40& 0.13 &  6.61 & 23& 409& H & T  &   \\
08416--2525& 08416--2525& 5.67&  3.92&  2.70&  1.38& 0.30 &  5.92 &  2& 454& A & EE &   \\
V875 Cen  & 11514--5841 &     &      &      &  5.13& 2.39 &  7.46 & 19& 599& B & E  & X \\
C3268     & 12374--5706 & 5.70&  4.42&  3.52&  2.73& 1.38 &  6.87 &  2& 398& A & SM &   \\
15148--4940& 15148--4940& 6.68&  4.18&  2.43&  0.34& 1.02 &  5.17 &  3& 590& A & MEE&   \\
V CrB     & 15477+3943 &  3.42&  2.07&  1.27&  0.32& 0.04 &  4.71 & 48& 358& H & T  &   \\
NP Her    & 16150+2558 &  5.67&  4.32&  3.52&  2.82& 0.17 &  7.06 &  2& 448&   & SM &   \\
T Dra     & 17556+5813 &  4.54&  2.78&  1.58&  0.22& 0.12 &  4.83 &  2& 422&   & MN &   \\
U Lyr     & 19184+3746 &  4.91&  3.42&  2.55&  1.62& 0.39 &  6.02 & 11& 452&   & T  &   \\
V1421 Aql & 19248+0658 &  7.41&  5.24&  3.71&  1.91& 1.07 &  6.38 & 12& 493&B  & E  & X \\
V1991 Cyg & 19419+3222 &  5.26&  3.98&  3.07&      & 1.01 &  6.39 &  1& 562&D  & M  & X \\
V359  Vul & 19537+2212 &  6.93&  4.78&  3.03&      & 0.96 &  5.87 &  1& 478&D  & M  & X \\
KL Cyg    & 19559+3301 &  5.37&  3.72&  2.70&      & 1.01 &  6.08 &  1& 526&   & M  & X \\
V1968 Cyg & 19594+4047 & 11.25&  7.53&  4.74&      & 0.81 &  5.37 &  1& 783&J  & JM &   \\
V2004 Cyg & 20004+2943 &  8.66&  6.14&  4.35&      & 1.95 &  7.12 &  1& 585&D  & M  & X \\
V1969 Cyg & 20072+3116 &      &      &  3.69&  1.23& 0.92 &  5.71 &  5& 550&J  & J  & X \\
WX Cyg    & 20166+3717 &  4.40&  3.22&  2.36&  2.08& 0.88 &  5.84 &  1& 399&H  & N  & X \\
U Cyg     & 20180+4744 &  3.33&  1.95&  1.16&  0.43& 0.62 &  4.60 &  7& 460&H  & T  &   \\
BD Vul    & 20351+2618 &  5.27&  3.89&  3.11&  2.43& 0.60 &  6.61 & 11& 430&   & ST &   \\
V Cyg     & 20396+4757 &  3.16&  1.19&  0.05&      & 0.40 &  3.41 &  1& 417&H  & M  & X \\
V703 Cep  & 20532+5554 & 13.73& 10.20&  7.40&      & 0.95 &  7.43 &  1& 591&N  & M  &   \\
V1549 Cyg & 21035+5136 &      &      &  2.91&  1.03& 0.71 &  5.54 &  9& 533&J  & J  & X \\
V573 Cyg  & 21070+4711 &  6.35&  4.67&  3.69&      & 1.47 &  6.98 &  1& 538&   & M  & X \\
AX Cep    & 21262+7000 &  4.04&  2.47&  1.54&      & 0.34 &  5.05 &  1& 395&   & M  &   \\
V1426 Cyg & 21320+3850 &  3.51&  1.85&  0.75&--0.48& 0.42 &  4.11 &  5& 470&   & TN & X \\
S Cep     & 21358+7823 &  2.41&  0.93&--0.07&      & 0.40 &  3.39 &  1& 486&   & M  &   \\
V1568 Cyg & 21366+4529 &  8.31&  6.19&  4.61&      & 1.01 &  7.32 &  1& 495&   & M  &   \\
V2345 Cyg & 21377+5042 & 10.83&  7.86&  5.67&      & 1.20 &  6.78 &  1& 500&N  & M  & X \\
V2358 Cyg & 21449+4950 & 12.54&  9.23&  6.75&      & 1.44 &  7.61 &  1& 646&N  & M  & X \\
V385 Lac  & 22236+5002 &  8.96&  6.60&  4.70&      & 0.83 &  7.35 &  1& 500&D  & M  &   \\
V384 Cep  & 22241+6005 &  9.05&  6.01&  3.98&      & 0.91 &  5.96 &  1& 698&N  & M  & X \\
MW Cep    & 22493+6018 &  5.91&  4.41&  3.76&      & 1.32 &  7.08 &  1& 400&   & M  & X \\
LL Peg    & 23166+1655 &      &      & 10.69&  3.87& 0.09 &  4.75 & 34& 696&B  &SEJE&  \\
LP And    & 23320+4316 &      &      &  2.71&  0.33& 0.34 &  4.40 & 11& 620&J  & J  &   \\
V955 Cas  & 23491+6243 & 12.68&  9.48&  7.00&      & 1.70 &  7.84 &  1& 575&N  & M  & X \\
\end{tabular}  
\end{center}
\flushleft 
{\bf $\dagger$ Source for periods if not GCVS:} {\bf A} Pojma\'{n}ski
(2003), Pojma\'{n}ski \& Maciejewski (2004); {\bf B} Le Bertre (1992); {\bf
C} mean of J and N; {\bf G} Groenewegen et al. (1998); {\bf J} Jones et al.
(1974); {\bf H} Perryman et al. (1997); {\bf N} Nakashima et al. (2000);
{\bf O} Olivier et al. (2001); {\bf D} Dahlmark (1993, 1996).
$\ddagger$ {\bf Source for near-infrared}: {\bf M} 2MASS, {\bf N}
Noguchi et al.(1981), {\bf T} Taranova \& Shenavrin (2004), {\bf E} ESO (see
section 4) {\bf J} Jones et al. (1990), {\bf S} SAAO from Paper I.
\end{table*}

\subsection{SAAO minus ESO.}
Le Bertre (1992) published extensive $JHKL$ photometry of a limited number
of C-Miras. Nine of these are in common with SAAO. In addition there are 
extensive, mostly single $JHKL$, ESO values for late-type stars including
some C-Miras in Epchtein et al. (1987, 1990), Fouqu\'{e} et al. (1992)
and Guglielmo et al. (1993). Altogether
there are 86 C-Miras in common between ESO and SAAO (84 in the case of $L$).
These yield: $\Delta J = +0.13 \pm 0.11$, $\Delta H = 0.00\pm 0.08$,
$\Delta K = +0.06 \pm0.07$, $\Delta L = +0.35 \pm 0.05$. The mean SAAO
$J-K$ for these stars is 4.15.  
These are the transformations we have adopted.
Bouchet et al. (1991) have discussed
the transformation between ESO and SAAO as a function of $(J-K)$, though
not for C-Miras. Their transformations, extrapolated to the mean
$(J-K)$ of the C-Mira sample give: $\Delta J = +0.09$, $\Delta H = -0.09$,
$\Delta K = +0.04$, $\Delta L = +0.01$. Within the uncertainties
the Bouchet transformations fit the C-Miras for $JHK$ but not at $L$.

\subsection {SAAO minus 2MASS.}
 Although the 2MASS data are generally for a single epoch only, there are 
sufficient stars (115) in common and a sufficient range in $(J-K)$
colour to divide the sample into three groups according to colour. There
is a trend in $J$ and $K$ but not in $H$. We find:
\begin{equation}
\Delta J = +0.051(J-K)_{2M} - 0.089
\end{equation}
\begin{equation}
\Delta H = -0.08
\end{equation}
\begin{equation}
\Delta K = -0.044(J-K)_{2M} +0.062.
\end{equation}
The uncertainty in these transformations is less than 0.10 mag.

\subsection {SAAO minus Taranova \& Shenavrin.}
 For northern observations (Taranova \& Shenavrin 2004 (TS), Jones et al. (1990),
Noguchi et al. (1981)) the direct overlap with SAAO observations is too
small to be useful. In these cases we have found the transformation to
the 2MASS system and then using the results just obtained have derived
the transformation to the SAAO system. There are 26 carbon-rich 
variables in common between 2MASS and Taranova \& Shenavrin, though
not all of these are Miras. This leaves out CW Leo which is much redder
than other stars in the sample. We then find for these stars (SAAO $-$ TS):
$\Delta J = -0.06$, $\Delta H = -0.18$,
$\Delta K = -0.11$. For a comparison between SAAO and TS at L we have only 
5 stars. However we also have the SAAO differences from the L of
Noguchi et al. (14 stars) as well as the Noguchi et al. and TS differences
(17 stars). These three differences evidently have to be consistent. We 
finally adopted $\Delta L = +0.10$.

\subsection {SAAO minus Noguchi et al..}
In the case of Noguchi et al. (1981)    
we have a direct comparison of $JHK$ with SAAO (14 stars), a comparison
via 2MASS (17 stars) using the SAAO -- 2MASS differences, and a comparison
via TS (17 stars)  using the differences found in the last section. 
In the mean we find: $\Delta J = +0.37$, $\Delta H = +0.10$,
$\Delta K = +0.10$. The value of $\Delta L$, derived directly from 14 stars
and indirectly as described in the previous paragraph is $+0.06$.
The relatively large difference at $J$ is perhaps not surprising in view of 
the narrow, non-standard, filter used (Okuda et al. 1974).

\subsection {SAAO minus Jones et al..}
    Jones et al. (1990) observed only at $K$ and $L$. The SAAO minus Jones
et al. difference at $K$ was derived from the Jones minus 2MASS difference
(22 stars) and using equation 8 above, at the relevant $(J-K)$. The $L$
difference comes from 6 star observed both by Jones and at SAAO. We find:
$\Delta K = -0.20$,
$\Delta L = +0.03$.

  All the results quoted in the above sections should be accurate to
0.10 mag or better. Table 3 lists the 
mean values of the non-SAAO photometry we have used,
converted to the SAAO system. The table lists the stars' name, IRAS number,
$JHKL$ data, $A_{V}$ and $m_{bol}$ calculated in the manner discussed in detail
in paper I and using the C-Mira PL($K$) relation discussed there with its LMC
zero-point, the number of near-infrared observations used, the period, 
a code letter for the source of the period if this was not taken
from the {\it General Catalogue of Variable Stars},
a code for the source of the photometry, 
and an indication of whether MSX data were
used.

\begin{center}
\onecolumn
\begin{longtable}{lrcrccrc}  
\caption[Distance and Velocity Data.]{Distance and Velocity Data.}\label{all}\\
\hline
& \multicolumn{2}{c}{ZP=1.87} & \multicolumn{2}{c}{ZP=2.06} \\
 name &\multicolumn{1}{c}{d}&\multicolumn{1}{c}{A$_V$} &
\multicolumn{1}{c}{d}&\multicolumn{1}{c}{A$_V$} & P &
\multicolumn{1}{c}{V$_{lsr}$} & note\\
& \multicolumn{1}{c}{(kpc)} & \multicolumn{1}{c}{(mag)}
& \multicolumn{1}{c}{(kpc)} & \multicolumn{1}{c}{(mag)} &
\multicolumn{1}{c}{(day)} & \multicolumn{1}{c}{($\rm km\,s^{-1}$)}\\
\hline
\endfirsthead

\hline
& \multicolumn{2}{c}{ZP=1.87} & \multicolumn{2}{c}{ZP=2.06} \\
 name &\multicolumn{1}{c}{d}&\multicolumn{1}{c}{A$_V$} &
\multicolumn{1}{c}{d}&\multicolumn{1}{c}{A$_V$} & P & \multicolumn{1}{c}{V$_{lsr}$} & note\\
& \multicolumn{1}{c}{(kpc)} & \multicolumn{1}{c}{(mag)}
& \multicolumn{1}{c}{(kpc)} & \multicolumn{1}{c}{(mag)} &
\multicolumn{1}{c}{(day)} & \multicolumn{1}{c}{($\rm km\,s^{-1}$)}\\
\hline
\endhead

  \multicolumn{8}{l}{{Continued on next page\ldots}} \\
\endfoot

  \\ \hline
\endlastfoot

\multicolumn{8}{l}{Stars with $JHKL$ photometry in Paper I }\\
YY Tri          & 2.25 & 0.24 & 2.06 & 0.24 & 624 & --3.4 & c \\
R For           & 0.70 & 0.04 & 0.64 & 0.04 & 385 & --1.0 & c \\
V718 Tau        & 1.46 & 1.35 & 1.34 & 1.34 & 388 &--16.2 & o \\
R Lep           & 0.47 & 0.25 & 0.43 & 0.24 & 438 & +16.7 & c \\
QS Ori          & 2.65 & 0.69 & 2.43 & 0.68 & 483 & +25.1 & c \\
05418--3224     & 2.73 & 0.04 & 2.51 & 0.04 & 483 & +27.6 & c \\
06088+1909      & 2.37 & 1.18 & 2.17 & 1.10 & 493 & +64.0 & c \\
ZZ Gem          & 1.76 & 0.61 & 1.62 & 0.58 & 316 & +66.8 & c \\
V617 Mon        & 2.94 & 1.01 & 2.69 & 0.96 & 444 &  +8.9 & o \\
V636 Mon        & 1.09 & 0.45 & 1.00 & 0.43 & 543 & +16.4 & c \\
V477 Mon        & 2.61 & 1.01 & 2.39 & 0.95 & 619 & +37.7 & c \\
RT Gem          & 3.43 & 0.54 & 3.14 & 0.54 & 350 &+101.1 & o \\
06487+0551      & 2.40 & 0.80 & 2.20 & 0.77 & 536 & +32.0 & c \\
CG Mon          & 1.99 & 0.75 & 1.82 & 0.71 & 424 & +14.8 & o \\
CL Mon          & 1.11 & 0.47 & 1.02 & 0.45 & 511 & +29.0 & c \\
06531--0216     & 2.01 & 0.86 & 1.85 & 0.80 & 595 & +39.4 & c \\
06564+0342      & 2.99 & 0.83 & 2.74 & 0.81 & 584 & +34.2 & c \\
07080--0106     & 3.76 & 0.16 & 3.44 & 0.16 & 594 & +42.5 & c \\
VX Gem          & 1.88 & 0.03 & 1.73 & 0.03 & 391 & +32.9 & o \\
R Vol           & 0.88 & 0.49 & 0.80 & 0.48 & 452 & --4.7 & c \\
07217--1246     & 2.18 & 0.84 & 2.00 & 0.80 & 620 & +24.9 & c \\
07220--2324     & 2.75 & 1.13 & 2.52 & 1.06 & 560 & +34.2 & c \\
$[$W71b$]$007--02&4.34 & 1.33 & 3.98 & 1.30 & 460 & +57.6 & o \\
07373--4021     & 1.09 & 0.54 & 1.00 & 0.51 & 459 & +24.3 & c \\
V471 Pup        & 4.17 & 1.43 & 3.83 & 1.39 & 390 & +90.5 & o \\
07454--7112     & 0.83 & 0.45 & 0.76 & 0.43 & 511 &--32.6 & c \\
V831 Mon        & 2.33 & 0.09 & 2.14 & 0.09 & 331 & +12.0 & o \\
07576--4054     & 2.86 & 1.25 & 2.62 & 1.20 & 519 &  +8.3 & c \\
07582--1933     & 2.49 & 0.44 & 2.28 & 0.43 & 541 & +10.2 & c \\
$[$ABC89$]$Pup38&12.30 & 0.98 &11.27 & 0.98 & 431 & +60.2 & o \\
FF Pup          & 3.77 & 0.58 & 3.46 & 0.57 & 431 & +48.3 & o \\
V518 Pup        & 2.03 & 0.10 & 1.86 & 0.10 & 448 &--17.9 & c \\
08050--2838     & 2.63 & 0.97 & 2.41 & 0.93 & 555 & --6.7 & c \\
08074--3615     & 2.40 & 1.52 & 2.20 & 1.45 & 832 & +14.4 & c \\
$[$ABC89$]$Ppx19& 5.89 & 2.19 & 5.39 & 2.18 & 474 & +40.4 & o \\
V346 Pup        & 1.36 & 0.88 & 1.25 & 0.83 & 568 &--16.6 & c \\
$[$ABC89$]$Ppx40& 5.13 & 1.54 & 4.70 & 1.53 & 428 & +64.8 & o \\
$[$W71b$]$029--02&4.88 & 3.04 & 4.47 & 2.98 & 470 & +59.0 & o \\
08340--3357     & 2.26 & 0.83 & 2.07 & 0.81 & 590 & +52.4 & c \\
R Pyx           & 1.35 & 0.38 & 1.23 & 0.37 & 369 & +25.4 & c \\
08535--4724     & 3.51 & 4.53 & 3.22 & 4.42 & 570 & +48.2 & c \\
08534--5055     & 4.29 & 2.20 & 3.93 & 2.11 & 703 & +12.2 & c \\
IQ Hya          & 1.55 & 0.53 & 1.42 & 0.52 & 382 &--12.7 & o \\
CQ Pyx          & 1.14 & 0.49 & 1.04 & 0.48 & 659 &  +4.3 & c \\
09176--5147     & 3.17 & 3.64 & 2.91 & 3.38 & 431 & +48.1 & c \\
$[$W71b$]$046--02&5.12 & 5.63 & 4.73 & 5.55 & 265 & +28.3 & o \\
$[$ABC89$]$Vel44& 4.18 & 3.34 & 3.85 & 3.22 & 413 & +40.6 & o \\
CW Leo          & 0.14 & 0.07 & 0.13 & 0.06 & 651 &--23.2 & c \\
09513--5324     & 1.79 & 1.87 & 1.64 & 1.72 & 630 & --5.0 & c \\
09533--6021     & 4.81 & 1.48 & 4.40 & 1.45 & 714 &  +9.8 & c \\
09521--7508     & 1.12 & 0.72 & 1.03 & 0.70 & 539 & --0.8 & c \\
10098--5742     & 3.91 & 3.54 & 3.59 & 3.17 & 585 &--31.2 & c \\
RW LMi          & 0.46 & 0.09 & 0.42 & 0.08 & 617 & --0.4 & c \\
CZ Hya          & 1.35 & 0.13 & 1.24 & 0.13 & 444 & +17.9 & c \\
TV Vel          & 1.84 & 1.08 & 1.69 & 1.01 & 404 &--10.2 & o \\
$[$ABC89$]$Car73& 6.05 & 4.52 & 5.59 & 4.19 & 483 &  +3.3 & o \\
$[$ABC89$]$Car84& 4.54 & 4.45 & 4.22 & 4.20 & 501 &  +0.4 & o \\
$[$ABC89$]$Car87& 7.09 & 3.46 & 6.58 & 3.28 & 473 & +20.4 & o \\
$[$ABC89$]$Car105&2.95 & 2.57 & 2.72 & 2.36 & 497 & --5.5 & o \\
11145--6534     & 1.73 & 1.51 & 1.58 & 1.41 & 623 &--20.7 & c \\
$[$W65$]$ c13   & 5.38 & 4.53 & 5.00 & 4.28 & 395 &--48.5 & o \\
$[$TI98$]$1130--1020&2.16&0.09& 1.98 & 0.09 & 443 & +23.0 & c \\
11318--7256     & 0.66 & 0.56 & 0.61 & 0.52 & 526 &  +1.1 & c \\
$[$ABC89$]$Cen4 & 4.00 & 1.68 & 3.69 & 1.57 & 514 &--14.4 & o \\
11463--6320     & 3.13 & 2.77 & 2.87 & 2.59 & 615 &  +6.2 & c \\
$[$ABC89$]$Cen32& 4.40 & 2.58 & 4.05 & 2.44 & 652 &--13.9 & o \\
$[$ABC89$]$Cen43& 6.06 & 3.93 & 5.59 & 3.74 & 535 & +23.1 & o \\
$[$ABC89$]$Cen60& 5.25 & 3.50 & 4.85 & 3.33 & 414 &--42.0 & o \\
CF Cru          & 5.88 & 5.80 & 5.51 & 5.51 & 430 &--16.7 & o \\
12194--6007     & 2.84 & 2.21 & 2.61 & 2.05 & 627 &--48.9 & c \\
12298--5754     & 1.85 & 1.58 & 1.70 & 1.48 & 580 &--26.0 & c \\
12394--4338     & 1.33 & 0.34 & 1.22 & 0.33 & 551 &--29.1 & c \\
RU Vir          & 0.91 & 0.08 & 0.84 & 0.08 & 444 &  +1.5 & c \\
V Cru           & 1.47 & 0.98 & 1.35 & 0.92 & 380 &--33.5 & o \\
12540--6845     & 1.37 & 0.65 & 1.25 & 0.61 & 586 &--32.2 & c \\
13343--5807     & 2.40 & 1.94 & 2.20 & 1.83 & 556 & --3.5 & c \\
13477--6532     & 2.32 & 1.36 & 2.12 & 1.26 & 690 &--41.5 & c \\
13482--6716     & 1.70 & 0.87 & 1.56 & 0.81 & 500 &--33.5 & c \\
13509--6348     & 2.98 & 2.30 & 2.73 & 2.10 & 678 &--26.6 & c \\
$[$ABC89$]$Cir26& 3.81 & 4.11 & 3.52 & 3.73 & 495 &--49.1 & o \\
$[$ABC89$]$Cir27& 3.57 & 4.14 & 3.29 & 3.83 & 538 &--21.7 & c \\
14395--5656     & 6.58 & 3.41 & 6.05 & 3.33 & 488 &--74.3 & o \\
14404--6320     & 3.62 & 2.55 & 3.32 & 2.42 & 643 &--73.9 & c \\
14443--5708     & 5.17 & 3.43 & 4.74 & 3.24 & 723 &--58.1 & c \\
15082--4808     & 0.95 & 0.60 & 0.87 & 0.56 & 632 & --1.6 & c \\
15084--5702     & 3.48 & 3.58 & 3.19 & 3.15 & 948 &--42.3 & c \\
II Lup          & 0.64 & 0.48 & 0.59 & 0.44 & 576 &--13.6 & c \\
16079--4812     & 2.23 & 3.50 & 2.05 & 3.18 & 710 &--42.2 & c \\
16171--4759     & 2.76 & 2.57 & 2.54 & 2.30 & 560 & +39.7 & c \\
V Oph           & 0.78 & 0.90 & 0.71 & 0.88 & 294 &--28.2 & c \\
17047--2848     & 3.22 & 1.31 & 2.95 & 1.29 & 531 & --6.8 & c \\
V2548 Oph       & 1.09 & 0.87 & 0.99 & 0.82 & 747 & --5.1 & c \\
SZ Ara          & 2.44 & 0.47 & 2.23 & 0.47 & 222 &--13.6 & o \\
V617 Sco        & 1.30 & 1.19 & 1.19 & 1.07 & 524 & --4.5 & o \\
17217--3916     & 2.59 & 1.71 & 2.37 & 1.53 & 630 & --6.4 & c \\
17222--2328     & 2.64 & 2.58 & 2.42 & 2.52 & 603 &--61.3 & c \\
V833 Her        & 1.07 & 0.14 & 0.98 & 0.14 & 540 & --7.2 & c \\
17446--4048     & 1.40 & 0.80 & 1.29 & 0.74 & 545 &  +1.5 & c \\
17581--1744     & 2.43 & 1.78 & 2.24 & 1.62 & 628 & +23.0 & c \\
18036--2344     & 2.13 & 2.42 & 1.95 & 2.18 & 664 & +21.3 & c \\
FX Ser          & 1.13 & 1.39 & 1.04 & 1.28 & 519 & +27.5 & c \\
V1280 Sgr       & 1.35 & 1.03 & 1.24 & 0.93 & 532 &  +0.6 & o \\
18119--2244     & 2.39 & 1.83 & 2.19 & 1.64 & 611 & +21.2 & c \\
V5104 Sgr       & 1.05 & 0.63 & 0.96 & 0.58 & 655 & +46.4 & c \\
18239--0655     & 1.65 & 1.53 & 1.51 & 1.38 & 635 & --2.8 & c \\
V1076 Her       & 1.14 & 0.36 & 1.04 & 0.35 & 609 & +56.2 & c \\
18248--0839     & 2.13 & 1.41 & 1.95 & 1.26 & 659 & +28.3 & c \\
V1417 Aql       & 0.87 & 0.36 & 0.80 & 0.32 & 617 & --0.1 & c \\
V821 Her        & 0.75 & 0.69 & 0.69 & 0.64 & 524 & --3.8 & c \\
AI Sct          & 2.16 & 1.07 & 1.99 & 0.97 & 408 & +23.0 & o \\
V1418 Aql       & 1.04 & 0.77 & 0.95 & 0.69 & 562 & +10.3 & c \\
V1420 Aql       & 1.00 & 0.52 & 0.92 & 0.49 & 694 & +17.9 & c \\
V1965 Cyg       & 1.13 & 0.56 & 1.03 & 0.51 & 577 &--16.2 & c \\
R Cap           & 1.62 & 0.46 & 1.48 & 0.46 & 349 &  +0.1 & c \\
V442 Vul        & 1.41 & 0.58 & 1.29 & 0.57 & 661 & --3.3 & c \\
RV Aqr          & 0.75 & 0.18 & 0.69 & 0.18 & 433 & --2.6 & c \\
IZ Peg          & 1.70 & 0.19 & 1.56 & 0.19 & 486 & +43.9 & c \\
\multicolumn{8}{l}{Stars with $JHKL$ photometry in Table 3 }\\                                                   
V668 Cas        & 1.56 & 0.63 & 1.43 & 0.59 & 650 &--30.9 & c \\
W Cas           & 1.52 & 0.67 & 1.40 & 0.63 & 399 &--44.3 & o \\
HV Cas          & 1.49 & 0.55 & 1.37 & 0.52 & 527 &--22.7 & c \\
PT Cas          & 3.98 & 0.75 & 3.64 & 0.75 & 300 &--90.1 & o \\
X Cas           & 1.37 & 0.64 & 1.26 & 0.60 & 420 &--54.0 & o \\
V596 Per        & 2.49 & 1.06 & 2.28 & 1.01 & 815 &  +4.9 & c \\
V409 Per        & 1.87 & 0.50 & 1.71 & 0.49 & 355 &--40.9 & o \\
V701 Cas        & 1.72 & 0.52 & 1.58 & 0.50 & 567 &--18.0 & c \\
KX Cam          & 3.01 & 1.73 & 2.76 & 1.69 & 430 &--54.6 & c \\
V384 Per        & 0.74 & 0.36 & 0.67 & 0.33 & 537 &--17.2 & c \\
KY Cam          & 2.31 & 1.10 & 2.12 & 1.02 & 477 &--87.7 & c \\
Y Per           & 1.58 & 0.49 & 1.45 & 0.47 & 238 & --8.1 & c \\
V414 Per        & 1.19 & 0.46 & 1.09 & 0.43 & 515 &--23.6 & c \\
V676 Per        & 3.49 & 0.74 & 3.20 & 0.74 & 482 &--48.7 & c \\
AU Aur          & 1.60 & 0.67 & 1.47 & 0.63 & 377 &--14.7 & o \\
DY Aur          & 1.85 & 0.82 & 1.70 & 0.76 & 474 & +18.3 & o \\
OP Aur          & 4.11 & 1.42 & 3.77 & 1.39 & 500 &  +9.0 & o \\
V370 Aur        & 1.90 & 0.82 & 1.74 & 0.76 & 732 &--30.0 & c \\
V393 Aur        & 2.02 & 0.57 & 1.85 & 0.55 & 517 &--45.7 & c \\
AZ Aur          & 1.63 & 0.54 & 1.50 & 0.51 & 416 & +71.4 & o \\
V1259 Ori       & 1.58 & 0.60 & 1.45 & 0.57 & 696 & +45.6 & c \\
V Aur           & 1.70 & 0.34 & 1.56 & 0.34 & 349 &  +0.3 & c \\
V713 Mon        & 2.21 & 0.65 & 2.02 & 0.62 & 494 & +28.4 & c \\
V688 Mon        & 1.77 & 0.90 & 1.63 & 0.81 & 653 &  +4.0 & c \\
HX CMa          & 1.69 & 0.74 & 1.55 & 0.70 & 725 & +13.0 & c \\
T Lyn           & 1.88 & 0.13 & 1.72 & 0.13 & 409 & --3.1 & o \\
08416--2525     & 1.45 & 0.30 & 1.33 & 0.29 & 454 &--13.7 & c \\
V875 Cen        & 3.09 & 2.39 & 2.83 & 2.30 & 599 &--13.7 & c \\
C3268           & 2.10 & 1.38 & 1.93 & 1.31 & 398 &--42.0 & o \\
15148--4940     & 1.17 & 1.02 & 1.07 & 0.95 & 590 &--41.6 & c \\
V CrB           & 0.73 & 0.04 & 0.67 & 0.04 & 358 &--102.5 & c \\
NP Her          & 2.42 & 0.17 & 2.22 & 0.17 & 448 &  +5.7 & o \\
T Dra           & 0.84 & 0.12 & 0.77 & 0.12 & 422 &--17.5 & c \\
U Lyr           & 1.51 & 0.39 & 1.38 & 0.38 & 452 &  +2.9 & o \\
V1421 Aql       & 1.86 & 1.07 & 1.71 & 0.99 & 493 & --5.8 & c \\
V1991 Cyg       & 2.00 & 1.01 & 1.84 & 0.96 & 562 & +16.7 & c \\
V359  Vul       & 1.45 & 0.96 & 1.33 & 0.88 & 478 & +19.8 & c \\
KL Cyg          & 1.68 & 1.01 & 1.54 & 0.93 & 526 & +33.7 & c \\
V1968 Cyg       & 1.48 & 0.81 & 1.36 & 0.76 & 783 & +25.7 & c \\
V2004 Cyg       & 2.85 & 1.95 & 2.62 & 1.78 & 585 & +13.7 & c \\
V1969 Cyg       & 1.44 & 0.92 & 1.32 & 0.84 & 550 & +13.6 & c \\
WX Cyg          & 1.30 & 0.88 & 1.20 & 0.81 & 399 & +46.3 & o \\
U Cyg           & 0.79 & 0.62 & 0.73 & 0.59 & 460 & +19.8 & c \\
BD Vul          & 1.93 & 0.60 & 1.77 & 0.58 & 430 & +39.5 & o \\
V Cyg           & 0.44 & 0.40 & 0.40 & 0.37 & 417 &  +9.8 & c \\
V703 Cep        & 3.31 & 0.95 & 3.04 & 0.91 & 591 & --2.0 & c \\
V1549 Cyg       & 1.31 & 0.71 & 1.20 & 0.66 & 533 &  +2.9 & c \\
V573 Cyg        & 2.57 & 1.47 & 2.36 & 1.37 & 538 &--26.2 & c \\
AX Cep          & 0.90 & 0.34 & 0.83 & 0.33 & 395 &  +9.5 & c \\
V1426 Cyg       & 0.64 & 0.42 & 0.59 & 0.40 & 470 & --9.3 & c \\
S Cep           & 0.47 & 0.40 & 0.43 & 0.38 & 486 &--19.1 & c \\
V1568 Cyg       & 2.88 & 1.01 & 2.64 & 0.98 & 495 & +27.8 & c \\
V2345 Cyg       & 2.24 & 1.20 & 2.06 & 1.10 & 500 &--12.1 & c \\
V2358 Cyg       & 3.68 & 1.44 & 3.38 & 1.39 & 646 &--34.1 & c \\
V385 Lac        & 2.94 & 0.83 & 2.69 & 0.82 & 500 &--35.9 & c \\
V384 Cep        & 1.83 & 0.91 & 1.68 & 0.84 & 698 & --9.7 & c \\
MW Cep          & 2.31 & 1.32 & 2.13 & 1.18 & 400 &--10.6 & c \\
LL Peg          & 1.05 & 0.09 & 0.96 & 0.09 & 696 &--34.3 & c \\
LP And          & 0.84 & 0.34 & 0.77 & 0.33 & 620 &--20.5 & c \\
V955 Cas        & 3.94 & 1.70 & 3.61 & 1.66 & 575 & --7.2 & c \\
\end{longtable}
\end{center}
\twocolumn

\section{Adopted Radial velocities}
 In addition to our own radial velocities listed in Table 2 we
have searched the literature for velocities of C-Miras for which we have
periods and $m_{bol}$ values. The main references were mentioned in the
Introduction. Early work (including the pioneering work of Sanford) is
listed by Abt \& Biggs (1972). Later compilations (e.g. Duflot et al. 1995,
Malaroda et al. 2000, Barbier-Brossat \& Figon 2000) were also consulted for
optical velocities.  Additional data were taken from Zuckerman \& Dyke (1989)
and Metzger \& Schechter (1994).

In view of the discussion in section 3 we have given preference to the CO
velocity where it is available. Otherwise mean optical velocities were used
with the corrections discussed in section 3 and giving unit weight to each
measurement. This, for instance, gives high weight to the data of Barnbaum
(1992a) who generally had several observations per star.  Because of the
relatively high resolution she used, her velocities should be of high
internal accuracy. All the velocities were converted to the local standard
of rest which is discussed in 
Feast et al. (2006)(paper III) and are tabulated in
Table 4. This table is divided into two sections. The first lists
the stars for which SAAO photometry was used and the second, those for which
non-SAAO photometry from Table 3 was used. In the table we give the
heliocentric distances ($d$) and corresponding estimated visual interstellar
absorptions ($A_{V}$) derived as discussed in detail in paper I. The results
are given for two slightly different distance scales. Both depend on the
adoption of a PL($K$) relation. In the first case the zero-point is derived
from an adopted LMC modulus of 18.50 and in the second case the zero-point
derived in paper III is used. The table gives; the star name, the distances
and $A_{V}$ values just discussed, the period, the velocity with respect to
our adopted standard of rest and an indication (c or o) as to whether the
velocity was from CO mm observations or was an optical (or near-infrared)
determination.

\section{Emission-line Velocities}
  The optical spectra of C-Miras generally show emission lines at some
phases. H$\alpha$ is often particularly strong. This line may be important
in future work on the kinematics of distant (e.g. extragalactic) C-Miras
which are too faint to be observed in mm CO. For both O- and C-Miras
it is known that emission line velocities are offset from the absorption
line velocities. The present observations listed in Table 2 yield a
mean difference between the absorption and emission velocities (a -- e)
of:
\begin{equation}
a - e = +30.2 \pm 1.4\, \rm{km\,s^{-1}}
\end{equation}
for 32 stars of mean period
429 days.\\
For these data the emission-line velocity was derived solely from
H$\alpha$. On the other hand existing data in the literature gives
the distinctly different value of:
\begin{equation}
a - e = +19.1 \pm 1.4\, \rm{km\,s^{-1}}
\end{equation}
for 40 stars of mean period 
405 days.\\
These velocities depend partly on emission lines other than H$\alpha$
(mainly other Balmer lines). However, Sanford (1944) found:
\begin{equation}
a - e = +20.4 \pm 0.9\, \rm{km\,s^{-1}} 
\end{equation}
for 34 stars
with the emission almost always from H$\alpha$.\\
Sanford (1950) gives a similar 
result for a few well observed stars. He also states that there is perhaps
a slight tendency for (a -- e) to increase with increasing period.

We have only six stars with a -- e values in common with earlier work.
These are CL Mon, ZZ Gem, V831 Mon, IQ Hya,
R Lep and V Oph. The earlier 
data for these stars are from Sanford (1944). These give:
\begin{equation} 
(a-e) = \rm +22 \pm 3 \,\rm km\,s^{-1}
\end{equation}
For the same stars the SAAO
data give:
\begin{equation}
 (a-e) = \rm +37 \pm 3\, \rm km\,s^{-1}.
\end{equation}
 The difference 
between the two results
\begin{equation}
\rm SAAO - Sanford = +14\, km\,s^{-1}
\end{equation}
is similar to that given by
the mean results above, but of lower weight. 

The reason for the difference between the present results and earlier ones
is not entirely clear. The blue shift of H$\alpha$ is due at least in
part to the mutilation of the emission line by overlying material in
the outer parts of the atmosphere (note that in O-Miras H$\alpha$ is
strongly suppressed by this mechanism). One factor that might therefore
affect (a -- e) is the optical depth of the circumstellar shell which 
could be greater for our sample than for earlier ones. 
However, there is no significant difference in the mean values of
$(K-L)$ for the set of SAAO stars and for those with earlier (a -- e)
values.
Another possibility is that our observations are more
randomly distributed round the light cycle than early work, some at least
of which will have been concentrated to maximum light where the molecular
band strength in the atmosphere is at its weakest. 
An instrumental effect cannot be entirely ruled out since H$\alpha$ was
measured against an argon comparison spectrum whereas the absorption
line velocities were derived by comparison with standard stars (see
section 2 above). However, the most likely cause is systematic differences
in the way the ``centre" of the emission line was defined in view of its
mutilation by overlying absorption. This is particularly so in view of
the different recording media and mode of measurement. The earlier results
were almost all obtained using photographic plates measured visually.
Whatever the cause it
is clear that it is a rather significant effect and the use of H$\alpha$
emission to measure the systemic velocity of a C-Mira will require 
caution. For our present purposes (paper III) this is not a matter
of importance since none of the radial velocities we adopt depend on
emission line measures.

\section*{Acknowledgments}
 We are grateful to Professor T Nagata for sending us a copy of the paper by
Okuda et al. and to the referee Dr Jacco van Loon for useful comments.

\end{document}